\documentclass[11pt]{article}
\usepackage[a4paper,margin=1in]{geometry}
\usepackage{amsmath,amssymb}
\usepackage{graphicx}
\usepackage{booktabs}
\usepackage[numbers]{natbib}
\usepackage{hyperref}
\usepackage{graphicx}
\usepackage{float}    
\usepackage{placeins} 

\usepackage{subcaption}

\title{\textbf{Real Time Magnetic Field Line Tracing in the Magnetosphere via Adaptive Error Bounded Integration}}
\author{
Taylan Demir\\
Department of Mathematics, Ankara University\\
Ankara, Turkey\\
}

\begin{document}
\maketitle

\begin{abstract}
Field line tracing is one of the fundamental computational tools used in the study of the magnetosphere, which helps in many areas including footprint mapping, connectivity analysis and real-time visualisation. This note describes an implementation approach to error-bounded adaptive integration of the field line differential equation (ODE), where an embedded Runge-Kutta pair is used in conjunction with event detection, allowing robust localisation of footprint locations on spacecraft. The method is validated against an analytic dipole field model as well as an international geomagnetic reference field (IGRF)-based geomagnetic configuration.
\end{abstract}
\section{Introduction}
The magnetosphere of Earth is a complex, ever-changing environment of charged particles, or plasma, created by the interaction of the solar wind (a continuous flow of charged particles from the sun) and Earth’s magnetic field [1,2,3]. In this environment, magnetic field lines act as a three-dimensional structures, providing a way to connect the various regions of the magnetosphere with corresponding locations in the upper atmosphere (the ionosphere) and to use them in the research and operational diagnostics of the magnetosphere [3]. Therefore, the process of following a magnetic field line or creating an imaginary line tangentially to the direction of the field is frequently performed to locate the position of a footprint of a magnetosphere or produce a quick visual representation of the magnetosphere, among other functions. The International Geomagnetic Reference Field (IGRF) serves as a benchmark for modelling the Earth's internal geomagnetic field by studying its time-dependence and measuring changes over time [4,5,6]. For those conducting research into the effects that external currents have on the geomagnetic field through the interaction of the solar wind and the place on Earth are located in that region, empirical models of the magnetosphere, like the Tsyganenko-style configuration, help create a model of the geomagnetic field around an object, providing an additional means to estimate the geomagnetic field due to the current conditions of the Sun [7,8]. When choosing which magnetic field model to use for any study, the process of integrating ODE for many different initial conditions still remains and, as you can see for the mapping of magnetic field traces, which occurs over time, so the computation of those traces are a major bottleneck for the computation of traces. Thus, this paper focuses on optimising the bottleneck from an algorithm standpoint rather than the creation of an entirely new magnetic field model and proposes a lightweight, fast, method for obtaining the very fast, reliable, and cost-effective tracing of magnetic field traces in a way where the trace can also have a high degree of accuracy in terms of how close they are to the actual locations of the trace. This is accomplished by blending (i) the use of an embedded Runge-Kutta integrator with variable step sizing which is a standard technique for solving non-stiff ODE's [9,10,11] and (ii) the application of sophisticated event detection techniques, so as to locate accurately where the trace crosses the spherical boundary of the Earth (Earth radius), allowing you to obtain dependable ionospheric and ground-height maps of where the trace crosses the Earth. While the aesthetic aspect is one of the primary motivations for real-time tracing, it is not the only consideration. In many pipelines, tracing-based quantities must be calculated many times, and a number of researchers have worked on the development and implementation of algorithms for tracing magnetic field lines in real-time [11]. Our contributions are threefold:
We have prepared and developed a normalized form of the Field-Line ODE that allows for easy identification and correction of geometric errors along with clearly defined practical termination conditions. The integration process is based upon an adaptive system of calculating errors during integration. The localisation of events has been combined with the calculation of footprints, so that batch footprints can be determined. We have created straightforward and simple benchmarks (verification of the dipole field; illustration on IGRF fields) that describe how the footprints change when tolerance is refined, the number of function evaluations and the amount of time it took to calculate them.
The remainder of the paper is organized as follows. Section 2 presents the magnetic-field configurations and the field-line ODE, while Section 3 outlines the adaptive integrator and the event detection mechanism developed to solve those equations. Section 4 presents results from numerical computations, while Section 5 summarizes conclusions and discusses potential limitations and directions for future work.
\section{Problem Formulation}
\label{sec:formulation}

\subsection{Field-line ODE}
Let $\mathbf{B}(\mathbf{x},t)\in\mathbb{R}^3$ denote a prescribed geomagnetic field
at position $\mathbf{x}$ and epoch $t$. A magnetic field line is an integral curve
whose tangent is everywhere parallel to $\mathbf{B}$.
For numerical tracing we use the normalized field-line ODE
\begin{equation}
\frac{d\mathbf{x}}{ds}=\mathbf{v}(\mathbf{x},t):=\frac{\mathbf{B}(\mathbf{x},t)}{\|\mathbf{B}(\mathbf{x},t)\|},
\label{eq:fieldlineODE}
\end{equation}
where $s$ is a curve parameter (arc-length-like). The normalization enforces
$\bigl\|\tfrac{d\mathbf{x}}{ds}\bigr\|=1$, which makes geometric error control
transparent: local numerical error is directly interpretable as a spatial
deviation along the traced curve.

\subsection{Coordinates, normalization, and initial conditions}
We work in geocentric coordinates and measure lengths in units of the Earth
radius $R_E$. Define the nondimensional position $\tilde{\mathbf{x}}=\mathbf{x}/R_E$,
so that the spherical boundary used for footprint detection is simply
\begin{equation}
g(\mathbf{x})=\|\mathbf{x}\|-R_E=0
\quad \Longleftrightarrow \quad
\|\tilde{\mathbf{x}}\|-1=0.
\label{eq:earthBoundary}
\end{equation}
Initial conditions are chosen as a set of seed points $\mathbf{x}_0$ in the
near-Earth region (e.g., on shells $\|\mathbf{x}_0\|=r_0$ with $r_0/R_E$ in a
moderate range). For every point of interest on our model, we make two integrations to find both sets of conjugate solutions and their points at which they intersect with the boundary of the model. These two sets of conjugate solutions are called the model dependent footprints of our model at that specific point of interest. More information about the bases for this method and the general applications of this kind of connectivity mapping can be located in standard texts on space physics, [3].

\subsection{Magnetic field models used in this note}
The aim here is algorithmic rather than model-development. We therefore adopt a
minimal two-level setup:

\paragraph{(A) Analytic dipole (verification).}
For verification and sanity checks, we use the classical dipole field
\begin{equation}
\mathbf{B}_{\mathrm{dip}}(\mathbf{x})
=\frac{1}{\|\mathbf{x}\|^{3}}\Bigl(3(\mathbf{m}\cdot\hat{\mathbf{x}})\hat{\mathbf{x}}-\mathbf{m}\Bigr),
\qquad
\hat{\mathbf{x}}=\frac{\mathbf{x}}{\|\mathbf{x}\|},
\label{eq:dipole}
\end{equation}
where $\mathbf{m}$ is a constant dipole moment vector. Since the present work
compares numerical tracing strategies rather than absolute amplitudes, the global
scale factor of $\mathbf{B}_{\mathrm{dip}}$ is immaterial for \eqref{eq:fieldlineODE}.

\paragraph{(B) IGRF internal field (realistic baseline).}
To create a realistic (non-empirical) reference for the geomagnetic field inside the Earth, we utilize the International Geomagnetic Reference Field (IGRF), which represents the average magnetic field over time through the use of spherical harmonic coefficients that characterize the magnetic field during specific years (Epoch) and its changes (Secular Variation) \cite{Alken2021IGRF13,NCEI_IGRF,IGRF14_Zenodo}.
In geocentric spherical coordinates $(r,\theta,\phi)$, the IGRF main-field model
is typically expressed through a scalar potential $V$ of the form
\begin{equation}
V(r,\theta,\phi,t)=a\sum_{n=1}^{N}\left(\frac{a}{r}\right)^{n+1}
\sum_{m=0}^{n}\Bigl(g_n^m(t)\cos(m\phi)+h_n^m(t)\sin(m\phi)\Bigr)P_n^m(\cos\theta),
\label{eq:igrfPotential}
\end{equation}
where $a$ is the reference radius, $P_n^m$ are associated Legendre functions, and
$N$ is the adopted truncation degree (as specified by the IGRF generation).
The magnetic field is obtained as $\mathbf{B}_{\mathrm{IGRF}}=-\nabla V$, and then
converted to Cartesian components for integration in \eqref{eq:fieldlineODE}.
Throughout this short note we restrict attention to the internal field component;
including external-current contributions (e.g., empirical magnetospheric models)
is left as a natural extension.
\section{Method: Adaptive Error-Bounded Integration with Footprint Events}
\label{sec:method}

\subsection{Integrator choice}
To follow the path indicated by Equation (1), we use an ERK pair that has an automatic selection of the step-size. An embedded pair creates two approximations for each of the equations depending on which order is used in each case (higher or lower), and the difference between the two gives us a good estimate of local error for any non-stiff problem [9,14]. For this reason, we will use the Dormand-Prince type of 5(4) methodology as a specific example [9]; however, the procedures below can be applied to other tableaux as well. Let $\mathbf{x}_n \approx \mathbf{x}(s_n)$ and $h_n$ be the current step size in the
curve parameter $s$. An ERK step produces:
\[
\mathbf{x}_{n+1}^{(5)}=\Phi^{(5)}(\mathbf{x}_n,h_n),\qquad
\mathbf{x}_{n+1}^{(4)}=\Phi^{(4)}(\mathbf{x}_n,h_n),
\]
where the superscripts denote the formal orders. We take
$\mathbf{x}_{n+1} := \mathbf{x}_{n+1}^{(5)}$ on accepted steps.

\subsection{Error norm and step-size control}
Define the componentwise scaling
\begin{equation}
\mathbf{s}_n = \mathbf{a} + r\,\max\bigl(|\mathbf{x}_n|,|\mathbf{x}_{n+1}^{(5)}|\bigr),
\label{eq:scaling}
\end{equation}
where $\mathbf{a}=(a,a,a)$ is an absolute tolerance vector and $r$ is a relative
tolerance. The embedded error estimate is
$\mathbf{e}_{n+1}=\mathbf{x}_{n+1}^{(5)}-\mathbf{x}_{n+1}^{(4)}$.
We measure it by the weighted RMS norm
\begin{equation}
\mathrm{err}_{n+1}=
\left(\frac{1}{3}\sum_{i=1}^{3}\left(\frac{e_{n+1,i}}{s_{n,i}}\right)^2\right)^{1/2}.
\label{eq:errnorm}
\end{equation}
A step is \emph{accepted} if $\mathrm{err}_{n+1}\le 1$ and \emph{rejected} otherwise.
Upon acceptance, we update the step size using the standard controller
\begin{equation}
h_{n+1} = h_n\,
\min\!\Bigl(\gamma_{\max},\max\bigl(\gamma_{\min},
\eta\,\mathrm{err}_{n+1}^{-1/(p+1)}\bigr)\Bigr),
\label{eq:stepsizeupdate}
\end{equation}
where $p$ is the order of the accepted solution (here $p=5$),
$\eta\in(0,1)$ is a safety factor (e.g.\ $\eta=0.9$), and
$0<\gamma_{\min}<1<\gamma_{\max}$ bound abrupt changes.
This accept/reject mechanism is classical for adaptive RK solvers [10,14].

\subsection{Stopping and safety criteria}
Because the velocity field $\mathbf{v}(\mathbf{x},t)=\mathbf{B}/\|\mathbf{B}\|$
in \eqref{eq:fieldlineODE} is undefined when $\|\mathbf{B}\|$ is extremely small,
we introduce a floor parameter $B_{\min}>0$ and terminate if $\|\mathbf{B}\|<B_{\min}$.
In addition, we stop integration when any of the following occurs:
(i) the trace reaches the footprint boundary (Section~\ref{sec:event}),
(ii) a prescribed maximal arclength $s_{\max}$ is exceeded,
or (iii) the trajectory exits a user-defined spatial domain
(e.g.\ $\|\mathbf{x}\|>r_{\max}$). These criteria prevent spurious long excursions
in regions where the chosen field model is not intended to be interpreted.

\subsection{Footprint event detection and localization}
\label{sec:event}
To compute a footprint on the spherical boundary $\|\mathbf{x}\|=R_E$, we define
the event function
\begin{equation}
g(\mathbf{x})=\|\mathbf{x}\|-R_E.
\label{eq:eventg}
\end{equation}
Assume an accepted step has produced $\mathbf{x}_{n+1}$. If
$g(\mathbf{x}_n)\,g(\mathbf{x}_{n+1})\le 0$, then the step brackets a crossing.
We then localize the root along the segment
\begin{equation}
\mathbf{x}(\theta)=\mathbf{x}_n+\theta(\mathbf{x}_{n+1}-\mathbf{x}_n),
\qquad \theta\in[0,1],
\label{eq:segment}
\end{equation}
by bisection (or a safeguarded secant method) until
$|g(\mathbf{x}(\theta^\ast))|\le \varepsilon_g$.
This produces a footprint point $\mathbf{x}_{\mathrm{fp}}=\mathbf{x}(\theta^\ast)$
with a user-set geometric tolerance $\varepsilon_g$.
Event handling in adaptive ODE contexts is a well-established practice; see,
for instance, discussions of event location strategies in practical ODE suites [10,14,15].

\subsection{Algorithm summary and computational cost}
We trace (1) from each seed point $\mathbf{x}_0$ in two opposite directions with respect to $\mathbf{s}$ (i.e., forwards and backwards), creating conjugate branches and footprints for them. The computational cost of doing this is primarily determined by how many times $\mathbf{B}$ is evaluated. For our 5(4) pair of Dormand–Prince, the total number of evaluations of $\mathbf{B}$ required to create each accepted step is fixed because they come from a fixed number of stages, whereas every rejected step adds to the cost by requiring an extra evaluation of $\mathbf{B}$. Thus, we summarize the total cost for each field line by the total number of evaluations of $\mathbf{B}$ and the wall time of the process, both of which we detail in Section 4.\\
\medskip
\noindent\textbf{Algorithm 1 (Adaptive field-line tracing with footprint event).}
\begin{enumerate}
\item \textbf{Input:} seed point $\mathbf{x}_0$, epoch $t$, tolerances $(a,r)$,
      initial step $h_0$, bounds $(s_{\max},r_{\max},B_{\min})$, and event tolerance
      $\varepsilon_g$.
\item Set $n=0$, $s_0=0$, $\mathbf{x}_0$ given, and choose direction $\sigma\in\{+1,-1\}$.
\item \textbf{While} not stopped:
  \begin{enumerate}
  \item Compute an embedded ERK step from $\mathbf{x}_n$ with step $\sigma h_n$,
        yielding $\mathbf{x}_{n+1}^{(5)}$ and $\mathbf{x}_{n+1}^{(4)}$.
  \item Form $\mathrm{err}_{n+1}$ by \eqref{eq:scaling}--\eqref{eq:errnorm}.
        \textbf{If} $\mathrm{err}_{n+1}>1$: reject step, decrease $h_n$ via
        \eqref{eq:stepsizeupdate}, and repeat.
  \item Accept: set $\mathbf{x}_{n+1}\leftarrow \mathbf{x}_{n+1}^{(5)}$,
        update $h_{n+1}$ via \eqref{eq:stepsizeupdate}, and advance $n\leftarrow n+1$.
  \item \textbf{Event test:} if $g(\mathbf{x}_{n-1})g(\mathbf{x}_{n})\le 0$ with
        $g$ in \eqref{eq:eventg}, localize $\mathbf{x}_{\mathrm{fp}}$ using
        \eqref{eq:segment} and stop.
  \item \textbf{Safety:} stop if $\|\mathbf{B}(\mathbf{x}_n,t)\|<B_{\min}$,
        if $|s_n|>s_{\max}$, or if $\|\mathbf{x}_n\|>r_{\max}$.
  \end{enumerate}
\item \textbf{Output:} traced polyline $\{\mathbf{x}_k\}_{k=0}^{n}$ and (if detected)
      footprint $\mathbf{x}_{\mathrm{fp}}$.
\end{enumerate}
\section{Numerical Experiments}
\label{sec:experiments}

This section validates the proposed adaptive tracing workflow (Section~\ref{sec:method})
on an analytic dipole field (verification) and then outlines a realistic baseline
configuration based on IGRF coefficients (demonstration). In all experiments we work
in nondimensional units with $R_E=1$, and we report both accuracy and cost metrics.

\subsection{Experiment A: Dipole verification}
\label{subsec:dipole}

We begin with the aligned dipole field \eqref{eq:dipole} (with $\mathbf{m}\parallel \hat{\mathbf{z}}$),
for which field-line geometry admits a closed-form characterization. In particular,
for a dipole line parameterized by magnetic latitude $\lambda$ one has the classical
relation
\begin{equation}
r(\lambda)=L\cos^2\lambda,
\label{eq:dipoleInvariant}
\end{equation}
where $L$ is constant along a field line (the so-called $L$-shell parameter)
[3]. For an equatorial seed at $r_0=L$ (i.e., $\lambda=0$),
the exact footprint latitude on $r=1$ is therefore
\begin{equation}
\lambda_{\mathrm{fp}}=\arccos\!\left(L^{-1/2}\right)=\arccos\!\left(r_0^{-1/2}\right).
\label{eq:dipoleExactFootprintLat}
\end{equation}
Since the tracing ODE is normalized, absolute scaling of $\mathbf{B}$ does not alter
the traced curve; only direction matters.

\paragraph{Setup.}
We choose seeds $\mathbf{x}_0=(r_0,0,0)$ with $r_0\in\{4,6,8\}$ and trace the ``south''
branch (integration direction aligned with the local field at the equator). Event
detection targets $g(\mathbf{x})=\|\mathbf{x}\|-1=0$ (Earth boundary). We test three
tolerance pairs $(r,a)$, corresponding to relative and absolute tolerances in the
weighted error norm \eqref{eq:errnorm}. The algorithm terminates at the first detected
boundary crossing.

\subsection{Accuracy metrics}
\label{subsec:metrics}

For dipole verification, an exact footprint $\mathbf{x}_{\mathrm{fp}}^{\ast}$ is
available via \eqref{eq:dipoleExactFootprintLat}. We measure footprint error by the
(chord) displacement on the unit sphere:
\begin{equation}
\Delta_{\mathrm{fp}} := \bigl\|\mathbf{x}_{\mathrm{fp}}-\mathbf{x}_{\mathrm{fp}}^{\ast}\bigr\|.
\label{eq:fpChord}
\end{equation}
As an interpretable angular proxy, we also report the absolute latitude error
$|\Delta\lambda| := |\lambda_{\mathrm{fp}}-\lambda_{\mathrm{fp}}^{\ast}|$ (degrees).

For realistic fields (e.g., IGRF), an analytic footprint is not available; in that
case we recommend a self-consistency metric using two tolerance levels:
\begin{equation}
\Delta_{\mathrm{fp}}^{(k)} := \bigl\|\mathbf{x}_{\mathrm{fp}}^{(k)}-\mathbf{x}_{\mathrm{fp}}^{(k+1)}\bigr\|,
\label{eq:selfConsistency}
\end{equation}
where $(k)$ indexes a tolerance refinement sequence. This directly quantifies how
stable the predicted footprint is under numerical refinement.

\subsection{Speed metrics}
\label{subsec:speed}
The major cost associated with tracing is analysing evaluation  $\mathbf{B}(\mathbf{x},t)$. Therefore, we present: (i) the number of field evaluations (nfev) and (ii) the wall time taken to trace a line (this is dependent on implementation and hardware and is useful for comparison purposes within this paper). Embedded Runge-Kutta (RK) Methods give a natural trade-off between accuracy cost and accuracy, as they provide for tighter tolerance resulting in more accepted steps and therefore less local error but require additional stages and, at times, one or more rejected steps [9,10,14].
\subsection{Results for Experiment A}
Table~\ref{tab:dipole} summarizes accuracy and cost for the dipole verification.
As expected, tightening tolerances decreases $\Delta_{\mathrm{fp}}$ and $|\Delta\lambda|$,
while increasing \texttt{nfev} (and typically wall time). Notably, a moderately strict
pair such as $(10^{-5},10^{-8})$ already achieves sub-millidegree latitude error across
the tested $L$-shells, while the loose setting $(10^{-2},10^{-5})$ can yield visibly
incorrect footprints for larger $r_0$.

\begin{table}[t]
\centering
\caption{Dipole verification: footprint accuracy and computational cost for seeds
$\mathbf{x}_0=(r_0,0,0)$ (south branch). $\Delta_{\mathrm{fp}}$ is the chord displacement
\eqref{eq:fpChord} relative to the exact dipole footprint from \eqref{eq:dipoleExactFootprintLat}.}
\label{tab:dipole}
\begin{tabular}{ccccccc}
\hline
$r_0/R_E$ & $(r,a)$ & \texttt{nfev} & steps & time [ms] & $\Delta_{\mathrm{fp}}$ & $|\Delta\lambda|$ [deg] \\
\hline
4 & $(10^{-2},10^{-5})$  & 56  & 7  & 4.07 & $2.18\times 10^{-2}$ & $1.25$ \\
4 & $(10^{-5},10^{-8})$  & 98  & 12 & 3.90 & $1.13\times 10^{-5}$ & $6.48\times 10^{-4}$ \\
4 & $(10^{-8},10^{-11})$ & 176 & 30 & 9.12 & $8.73\times 10^{-9}$ & $5\times 10^{-7}$ \\
\hline
6 & $(10^{-2},10^{-5})$  & 50  & 7  & 2.10 & $5.74\times 10^{-3}$ & $0.329$ \\
6 & $(10^{-5},10^{-8})$  & 110 & 13 & 6.62 & $1.30\times 10^{-5}$ & $7.46\times 10^{-4}$ \\
6 & $(10^{-8},10^{-11})$ & 206 & 35 & 7.43 & $1.07\times 10^{-8}$ & $6.14\times 10^{-7}$ \\
\hline
8 & $(10^{-2},10^{-5})$  & 68  & 8  & 4.13 & $1.00\times 10^{-1}$ & $5.75$ \\
8 & $(10^{-5},10^{-8})$  & 134 & 15 & 5.55 & $1.26\times 10^{-5}$ & $7.19\times 10^{-4}$ \\
8 & $(10^{-8},10^{-11})$ & 230 & 39 & 9.10 & $8.69\times 10^{-9}$ & $4.98\times 10^{-7}$ \\
\hline
\end{tabular}
\end{table}

\subsection{Experiment B: IGRF demonstration (baseline)}
\label{subsec:igrf}

For a realistic baseline, we use the International Geomagnetic Reference Field (IGRF)
coefficients and evaluate $\mathbf{B}_{\mathrm{IGRF}}=-\nabla V$ from the standard
spherical-harmonic potential representation \eqref{eq:igrfPotential} [4,5,6]. We fix a single epoch $t_0$ and select a seed set spanning a modest near-Earth shell (e.g., $\|\mathbf{x}_0\|=r_0$ with multiple
longitudes and colatitudes). For each seed we trace both branches and record the
first intersections with $\|\mathbf{x}\|=1$. The three principal products of the project include the following: (i) A footprint map represented as latitude and longitude on the surface of the earth, (ii) A 3-dimensional representation of the traces to allow a qualitative assessment of the traces created, and (iii) Performance statistics, which include an average/median number of \texttt{nfev} (number of function evaluations) per trace and per time unit traced. If it is not possible to obtain a closed-form 'exact' IGRF footprint, or if it is not reasonable to do so, then details of the performance under tolerance refinement as described in Equation (13) should be included. The observations suggest that once tolerances have become quite small, the practical benefit of further reducing the tolerances is not as great as when the tolerance was larger and, therefore, the operationally useful way to daily trace is to always work with the same tolerance.

%
\FloatBarrier
\begin{figure}[H]
\centering
\begin{subfigure}{0.98\linewidth}
  \centering
  \includegraphics[width=0.88\linewidth]{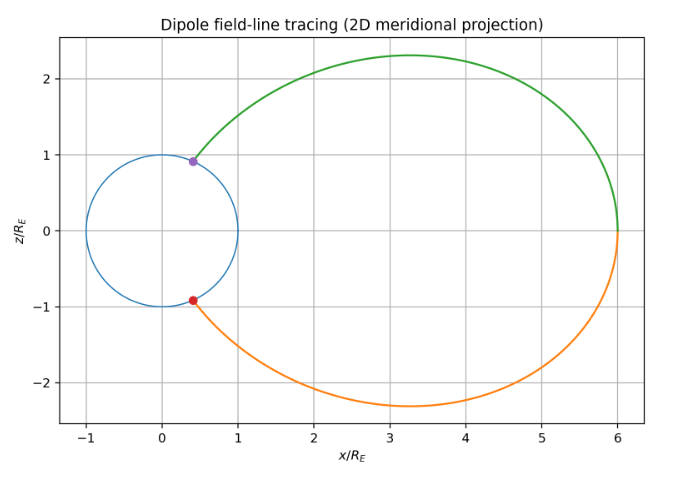}
  \caption{2D meridional projection and conjugate footprints.}
  \label{fig:dipole2D_sub}
\end{subfigure}

\vspace{0.6em}

\begin{subfigure}{0.98\linewidth}
  \centering
  \includegraphics[width=0.88\linewidth]{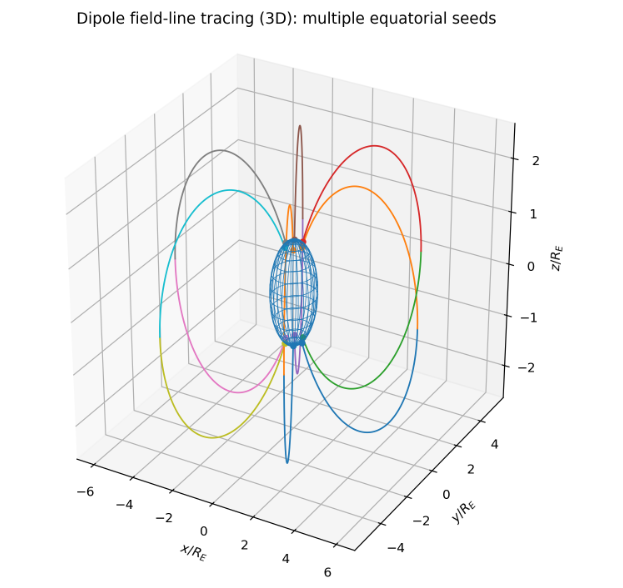}
  \caption{3D field lines for multiple equatorial seeds.}
  \label{fig:dipole3D_sub}
\end{subfigure}
\caption{Dipole field-line tracing validation in 2D and 3D.}
\label{fig:dipoleValidation}
\end{figure}
\FloatBarrier

\subsection{Limitations and future work: toward external-current coupling}
\label{subsec:future_external}
The current implementation of the method is, in part, limited by its representation of the geomagnetic field as an internal, dipole/IGRF type structure, as well as its exclusion of external current systems including ,for example, current systems associated with the magnetopause, the ring current, the tail current and Birkeland currents that are responsible for large-scale distortions in the topology of field lines and thus where and how these footprints will be located. At best, INF estimation of footprint location and connectivity will provide only estimates for quiet-time values since there will be additional contributions from all of these current systems that have not been included in the model or in the footprint locations calculated. The next immediate step will be to introduce an external field component into the footprint tracing pipeline using established empirical field models of the magnetosphere and/or data-assimilative corrections based on solar wind/IMF conditions from upstream solar wind sources and geomagnetic activity indices (i.e., Kp, Dst). From a computational perspective, adding an external component to the tracing pipeline can take advantage of the adaptive and multiresolution infrastructure developed for this large-scale modelling process (i.e., adaptive wavelet-Galerkin solvers and wavelet-regularized pipelines) [16,17,18]. In combination with the above improvements to the realism of the physical description of the field, the revised methodology will allow for the systematic quantification of uncertainty in footprint locations and connections concerning the levels of geomagnetic activity, as well as create the potential for near-real-time use of this methodology for operational purposes.

\section{Conclusion}
\label{sec:conclusion}
A compact computational pipeline for tracing geomagnetic field lines in idealized internal field configurations was displayed with consistent two and three dimensionalities and diagnostics that generate an expected dipole geometry and conjugate footprint structure. These resulting images contain a transparent validation layer for future exploration into numerical experiments and allow for reproducible comparisons across differing seed placements and numerical step size choices. There are many immediate avenues for extension. The first is to relax the internal field assumption(s) by coupling the tracer with external current systems such as ring currents, magnetopause/tail currents, and field-aligned currents e.g., through the use of established empirical magnetospheric models as represented within the Tsyganenko family of models, allowing for footprint and connectivity analysis based on activity. The second is regarding development of a solver that has the capacity to handle larger ensembles of seeds accelerated through vectorization and batched integration and/or to be utilized in conjunction with GPGPU-aware implementations. Recent experiences with adaptive multiresolution and wavelet-regularized computational pipelines have indicated a path forward for these scaling techniques [16,17,18].

\bibliographystyle{plainnat}

\end{document}